\documentclass[transmag]{IEEEtran}
\usepackage{latexsym}
\usepackage{acronym}
\usepackage{graphicx}
\usepackage{amsfonts,amssymb,amsmath}
\usepackage{xcolor}
\PassOptionsToPackage{hyphens}{url}\usepackage{hyperref}
\usepackage{url}

\def\BibTeX{{\rm B\kern-.05em{\sc i\kern-.025em b}\kern-.08em T\kern-.1667em\lower.7ex\hbox{E}\kern-.125emX}}

\input{AcronymList.tex}

\begin{document}

\title{Generalized Coordinated Multipoint Framework for 5G and Beyond}

\author{Muhammad~Sohaib~J.~Solaija,~\IEEEmembership{Student Member, IEEE},~Hanadi~Salman,~\IEEEmembership{Student Member, IEEE},\\~Abuu B.~Kihero,~Mehmet~Izzet~Sa\u{g}lam~and~H{\"u}seyin~Arslan,~\IEEEmembership{Fellow,~IEEE}
\thanks{This work was supported in part by the \ac{TUBITAK} under Grant No. 116E078. The work of Mehmet Izzet Sa\u{g}lam was supported by \ac{TUBITAK} under Grant No. 3171084.}
\thanks{Muhammad Sohaib J. Solaija, Hanadi Salman, Abuu B. Kihero and H{\"u}seyin~Arslan are with the Department of Electrical and Electronics Engineering, Istanbul Medipol University, Istanbul, 34810, Turkey (e-mails: \{muhammad.solaija,~hanadi.suleiman,~abuu.kihero\}@std.medipol.edu.tr,~huseyinarslan@medipol.edu.tr).}
\thanks{H{\"u}seyin~Arslan is also with the Department of Electrical Engineering, University of South Florida, Tampa, FL, 33620, USA}
\thanks{Mehmet Izzet Sa\u{g}lam is with the Department of R\&D, Turkcell Teknoloji Arastirma ve Gelistirme Anonim Sirketi, Istanbul, 34854 Turkey (e-mail: izzet.saglam@turkcell.com.tr)}

\thanks{This work has been submitted to the IEEE for possible publication. Copyright may be transferred without notice, after which this version may no longer be accessible.}}

\IEEEtitleabstractindextext{\begin{abstract}The characteristic feature of 5G is the diversity of its services for different user needs. However, the requirements for these services are competing in nature, which impresses the necessity of a coordinated and flexible network architecture. Although coordinated multipoint (CoMP) systems were primarily proposed to improve the cell edge performance in 4G, their collaborative nature can be leveraged to support the diverse requirements and enabling technologies of 5G and beyond networks. To this end, we propose generalization of CoMP to a proactive and efficient resource utilization framework capable of supporting different user requirements such as reliability, latency, throughput, and security while considering network constraints. This article elaborates on the multiple aspects, inputs, and outputs of the generalized CoMP (GCoMP) framework. Apart from user requirements, the GCoMP decision mechanism also considers the CoMP scenario and network architecture to decide upon outputs such as CoMP technique or appropriate coordinating clusters. To enable easier understanding of the concept, popular use cases, such as vehicle-to-everything (V2X) communication and eHealth, are studied. Additionally, interesting challenges and open areas in GCoMP are discussed.\end{abstract}

}

\maketitle
\section{Introduction} 
\IEEEPARstart{T}{he} \ac{4G} of mobile communication primarily focused on improving the data rates for users. While the \ac{5G} has catered to the enhancement of data rates under the \ac{eMBB} service, it has also expanded its vision to incorporate the increasing number of wireless devices and stringent reliability and latency requirements under the \ac{mMTC} and \ac{uRLLC} services, respectively. The latter two services are of particular importance for the fourth industrial revolution which is signified by unprecedented automation, driven with massive connectivity. If we talk quantitatively about \ac{5G} goals, it envisions a 1000 times increase in the volume of data and number of connected devices per km$^2$, 100-fold improvement in data rates, and reduction in latency by a factor of five \cite{series2015imt}. These demands are unlikely to be achieved without significant technological advancements. 
 
 Some of the well-established \ac{5G} enablers include \ac{mmWave} communications, small cells, massive \ac{MIMO} antenna systems, beamforming, full duplex systems and more flexible and adaptive \ac{PHY} and \ac{MAC} layer designs \cite{IEEE_specturm}. Furthermore, technologies like \ac{THz} communication, \acp{RIS}, integrated hybrid aerial-terrestrial-satellite networks, and artificial intelligence empowered communication systems are being explored and studied for beyond \ac{5G} scenarios \cite{saad2019vision}.

\begin{figure*}[t!]
     \centering
     \includegraphics[width=0.8\textwidth]{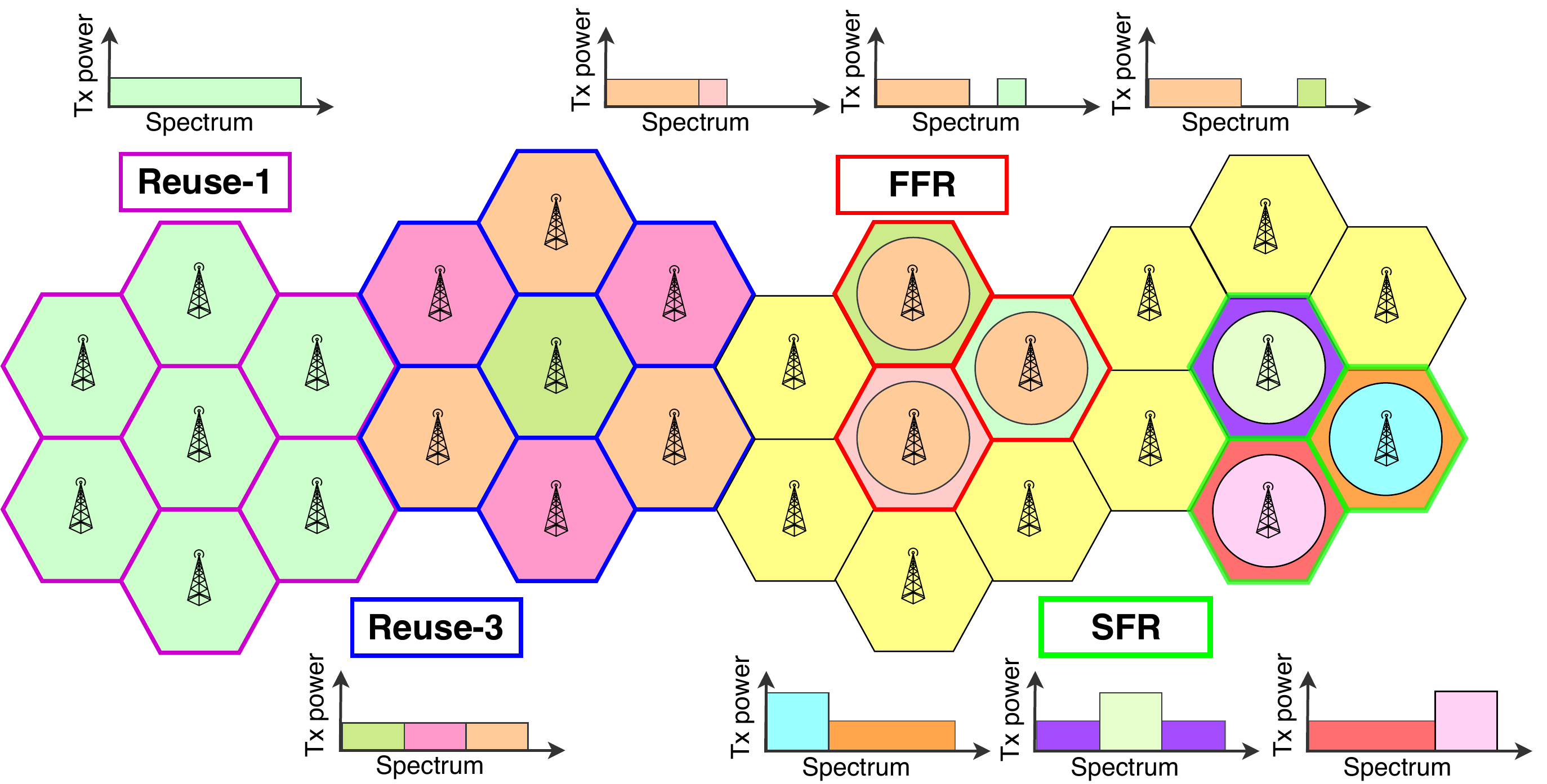}
     \caption{Different frequency reuse techniques for ICI avoidance} 
     \label{fig:ICIC techniques}
 \end{figure*}
 
Like the previous generations, \ac{5G} needs to address the increased throughput and capacity requirements of \ac{eMBB} which necessitates more frequent spectrum reuse, leading to smaller cells. This, combined with the envisioned 1000-fold increase in the number of connected devices, results in a significant increase in the number of \acp{UE} affected by \ac{ICI}. Traditional approaches for \ac{ICI} management have focused on minimizing it by increasing the reuse distance. However, these schemes are only capable of scheduling a cell's own resources without any knowledge of the neighboring ones. Access to such knowledge can help \acp{TP} schedule resources such that \ac{ICI} can be minimized across the network without wastage of resources. This led to the proposition of \ac{ICIC} and \ac{CoMP} transmission in the \ac{3GPP} Rel-8 and Rel-11, respectively. 

Moreover, the use of higher frequencies has made the utilization of beam-based transmission imperative in future networks. While these beams have lesser interference as compared to omnidirectional transmission at cell edges, they are much more prone to blockages. This makes meeting the reliability constraints for \ac{uRLLC} applications considerably more challenging. In such cases, coordinated transmission from different \acp{TP} is a possible solution. Although the primary motivation of \ac{CoMP} was to improve the performance of cell edge \acp{UE} by utilizing coordination between neighboring \acp{TP}, the spatial diversity provided by \ac{CoMP} can be exploited for addressing various \ac{5G} and beyond requirements. What we propose is to utilize \ac{CoMP} to satisfy the requirements such as reliability, throughput, latency, and \ac{PHY} layer security. Such a generalized realization of \ac{CoMP} requires a solid framework, with well-defined inputs, decision mechanisms, and outputs.

This paper attempts to address the above-mentioned gap in the present literature by contributing the following:
 
\begin{itemize}
    \item The idea for \ac{GCoMP} is presented and its associated framework is proposed, which takes into account the specific user requirements.
    \item Emerging technologies for future wireless networks - such as \ac{C-RAN}, \ac{RIS}, and hybrid networks - are discussed in relation to the \ac{GCoMP} framework.
    \item Specific use cases such as \ac{V2X} communication and eHealth are discussed under the \ac{GCoMP} framework.    
    \item Possible improvements and current challenges for the framework are discussed.
\end{itemize}

The rest of this article is structured in the following manner. First, a summary of the emergence of CoMP as an interference management technique is presented. Next, its potential as a solution to various conflicting goals of future networks is motivated, which is followed by a description of the proposed \ac{GCoMP} framework. To demonstrate the applicability of the framework, \ac{V2X} and eHealth use cases are considered as examples. Finally, different challenges that need to be faced in realizing such a framework are discussed.

\section{ICIC Evolution and \ac{CoMP} for \ac{5G}}
\label{sec:evolution}
This section follows the developments in \ac{ICIC} techniques, from the earlier generations of mobile communications to the proposition of CoMP in \ac{3GPP} Rel-11. Afterwards, projections are made regarding the compatibility of \ac{CoMP} and \ac{5G} technologies.

\subsection{ICIC Evolution}
Since early stages, the evolution of cellular systems has been hindered by the \ac{ICI} problem, which significantly degrades the performance of cell edge \acp{UE}. Early cellular technologies, like 2G, increased the frequency reuse distance to minimize the interference. Integer frequency reuse schemes, such as reuse-3 and 7 are primary examples of this approach. While this resulted in a significant reduction of interference, it also decreased the system capacity. Since the interference at the cell edge is the primary concern, solutions like \ac{FFR} were proposed, which split the cells into inner and outer regions and assigned the frequency bands such that inner regions have reuse-1 and outer regions have reuse-3 or reuse-7. A further extension to \ac{FFR} is \ac{SFR} which allocates higher power to the frequency bands allocated for cell edges. Such an approach improves the capacity of the system as compared to integer reuse, while still mitigating \ac{ICI} \cite{kosta2012interference}. Figure \ref{fig:ICIC techniques} illustrates the different schemes described above.

Despite the simplicity and effectiveness of the above-mentioned techniques, they are hampered by their standalone nature since there is no mechanism for different \acp{TP} to coordinate with each other for interference management. This led to the emergence of the \ac{ICIC} concept in \ac{3GPP} Rel-8. The basic working principle of \ac{ICIC} is that neighboring cells communicate with each other to figure out the best way to allocate resources for the cell edge \acp{UE} with minimum interference. \ac{ICIC} utilizes different flags, namely \ac{RNTP}, \ac{HII} and \ac{OI}, for coordination between \acp{TP} over the X2 interface \cite{kosta2012interference}. \Ac{RNTP} indicates to neighboring \acp{TP} the \acp{PRB} scheduled with high power. This enables the neighboring \acp{TP} to avoid assigning the same resources to their cell edge \acp{UE}, reducing the \ac{ICI}. \Ac{HII} is the uplink counterpart of \ac{RNTP}. On the other hand, \ac{OI} is a reactive function that indicates a high detected interference level on a specific resource block to neighbor \acp{TP}. When a cell scheduler receives \ac{OI}, it will change the scheduling to decrease the interference generated on these \acp{PRB}. 

Since \ac{ICIC} approaches focused on the data channels only, they allowed potential interference over control channels. This problem was compounded by the disparity in transmission powers in the case of \acp{HetNet}. Consequently, \ac{eICIC} was proposed where the different tiers use orthogonal resources to mitigate the interference. One example corresponding to this is the \ac{ABS} concept, where the small \ac{TP} only transmits during the \ac{TTI} where the macro \ac{TP} is muted. However, the rapid increase in device density, and with it the severity of the \ac{ICI} problem, necessitated more sophisticated and dynamic approaches. Consequently, \ac{CoMP} was introduced in the Rel-11 of \ac{3GPP} as a potential solution \cite{3GPP_36_819}.

\subsection{CoMP for \ac{LTE}}
\begin{figure}[t!]
    \centering
    \includegraphics[scale=0.67]{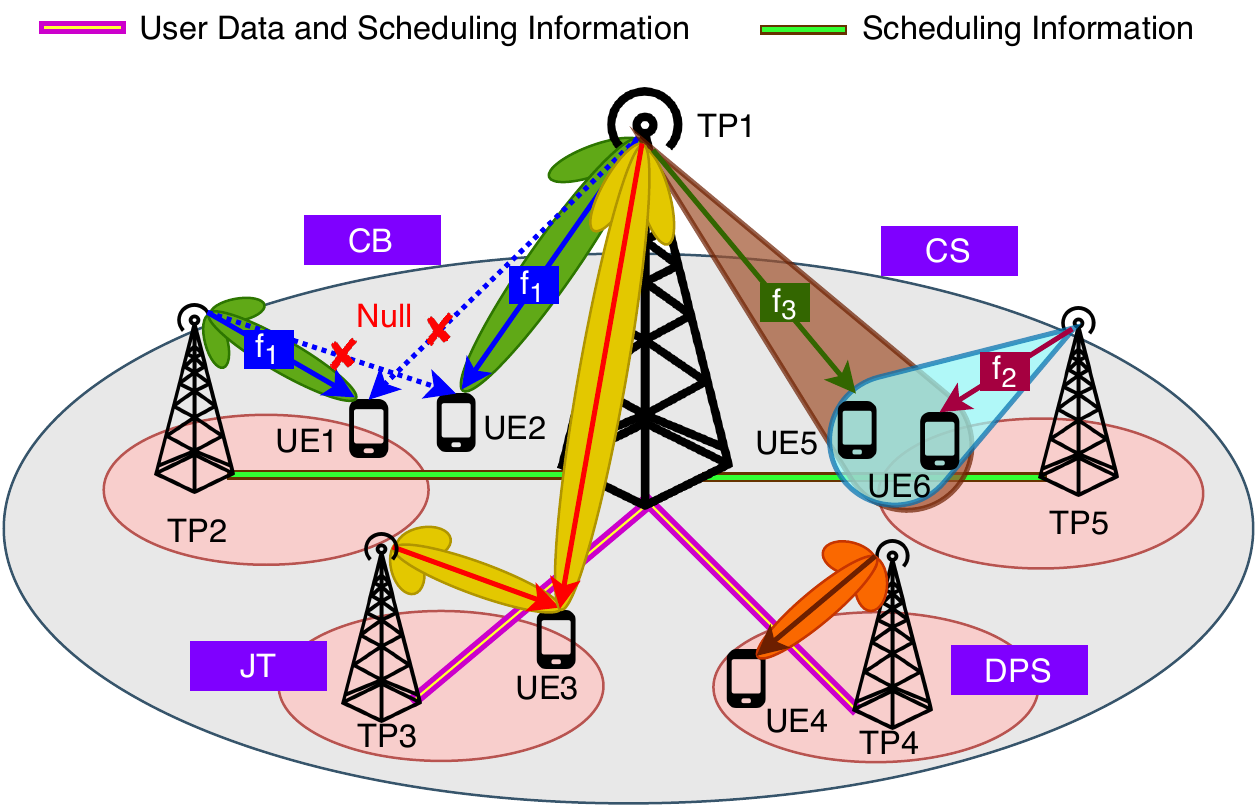}
    \caption{Illustration of CoMP schemes. CS = Coordinated Scheduling, CB = Coordinated Beamforming, JT = Joint Transmission, DPS = Dynamic Point Selection}
    \label{fig:CoMP Schemes}
\end{figure}

Unlike \ac{ICIC} approach and its variants (i.e., \ac{eICIC}) where \ac{ICI} is mitigated by restricting radio resource usage in either frequency or time domain, \ac{CoMP} incorporates spatial domain in resource allocation, which enhances spectral efficiency in addition to the interference mitigation. 
Figure \ref{fig:CoMP Schemes} illustrates the different \ac{CoMP} schemes. \Ac{CS}, which is essentially the evolution of previously mentioned interference mitigation techniques, reduces interference by ensuring instantaneous exchange of channel information between coordinating \acp{TP}. \Ac{CB} allows the edge \acp{UE} to use the same frequency resources as long as the beam patterns for different \acp{UE} do not interfere with each other. Due to the significant use of beamforming in \ac{5G} networks, \ac{CB} has attained increased importance. \Ac{JT}, arguably the most interesting \ac{CoMP} technique, constitutes of \ac{UE} data being transmitted from different \acp{TP}, potentially providing macrodiversity against path loss, shadowing, and blockage. \Ac{DPS} is a special case of \ac{JT}, where even though the \ac{UE} data is available at different \acp{TP}, it is only transmitted from one \ac{TP} at any given time \cite{qamar2017comprehensive}. Here, it is important to note that even though \ac{CoMP} mechanisms are devised to reduce the cell edge interference, they also improve the overall performance of the cell and network by more efficient resource allocation at the cell edges.

When the initial concept of \ac{CoMP} was introduced in Rel-11 \cite{3GPP_36_819} to improve the cell edge and system throughput, it was limited to \acp{TP} connected with ideal backhaul. It was not until Rel-12 that the concept was extended to multiple \acp{eNB} connected with non-ideal backhaul \cite{3GPP_36_874}. This required the standardization of signaling over X2 interface to enable exchange of \textit{\ac{CoMP} hypothesis set} and its associated \textit{benefit metric} between cooperating \acp{eNB}. In addition to these, \ac{RSRP} measurements were also used to validate these hypotheses. The sharing of this information with the neighboring \acp{eNB} helps improve the radio resource management \cite{roessler2015lte}.

Enhancements to inter-\ac{eNB} \ac{CoMP} were introduced in Rel-13 relate to the use of \ac{CSI} and \ac{eRNTP}. The latter is particularly useful to control the power allocation in a coordinated setting \cite{3GPP_36_300}. The strict requirements of \ac{JT} regarding synchronization and accurate \ac{CSI} necessitated exploration of other alternatives, leading to discussion around \ac{NC-JT} in Rel-14. The performance results indicated the suitability of \ac{NC-JT} and  \ac{CS}/\ac{CB} in low and high traffic load scenarios, respectively \cite{3GPP_36_741}. A study for creation and management of \ac{CoMP} sets based on network conditions was carried out under the umbrella of \acp{SON} in Rel-15 \cite{3GPP_36_742}. This study focused on monitoring two parameters, i.e., X2 characteristics and the spatio-temporal traffic variation.

\subsection{CoMP for 5G and Beyond}
\begin{figure}[t!]
    \centering
    \includegraphics[scale=0.45]{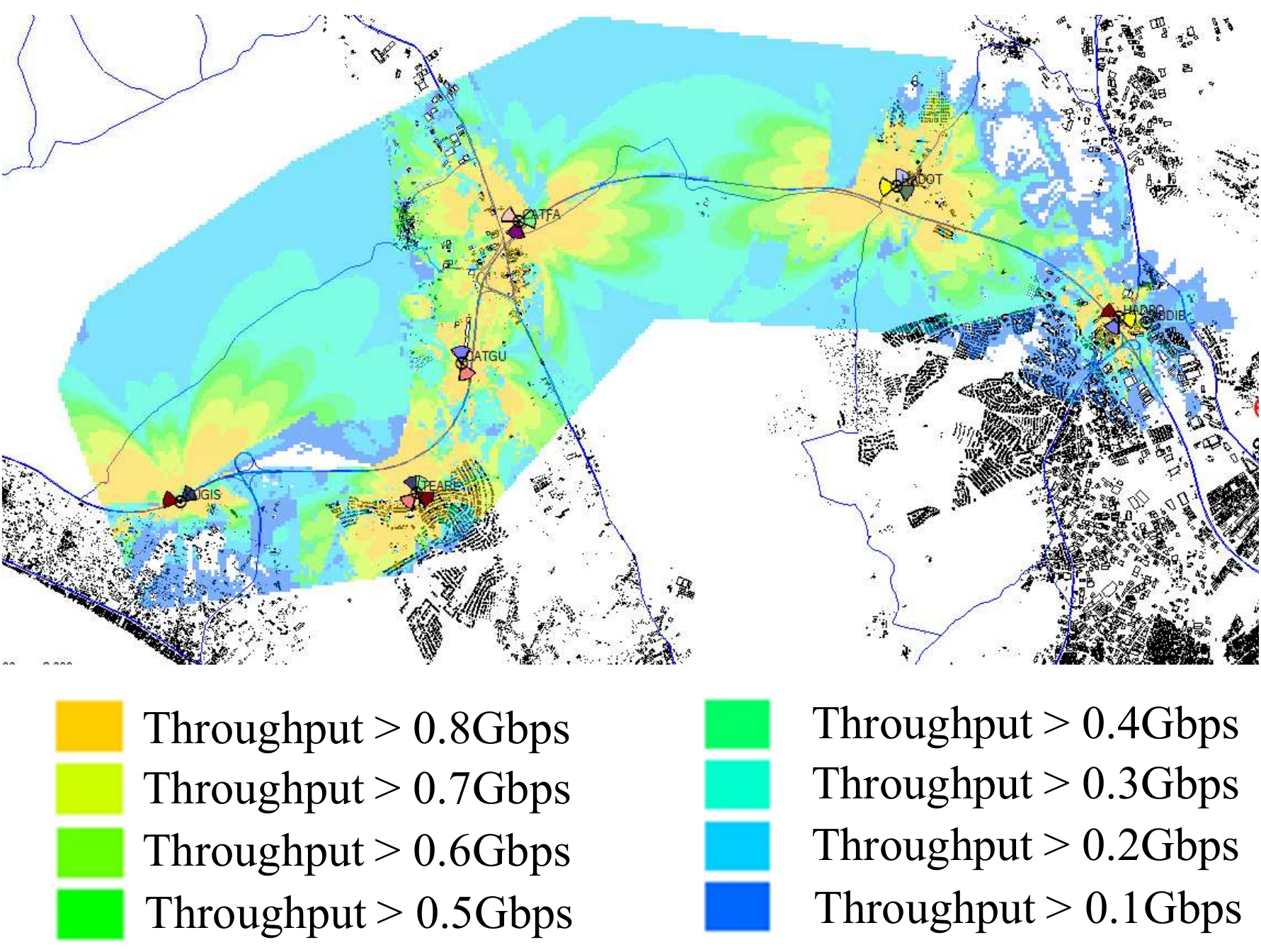}
    \caption{Coverage map of Istanbul \c{C}atalca Region - Turkey for different throughput requirements obtained using Atoll radio planning tool}
    \label{fig:Application Coverage}
\end{figure}

Unlike 4G, where the focus is primarily on improving the data rates, \ac{5G} and beyond networks have a myriad of requirements and objectives that need to be fulfilled. This requires a multitude of technologies to be introduced in the \ac{5G} and beyond networks. The following items look at the potential role of \ac{CoMP} for some of these mechanisms:

\begin{itemize}
    \item \ac{3GPP} Rel-14 specifies eight different functionality splits between central and distributed units for \ac{5G} \cite{3GPP_38_801}. The functional split has a major impact on the backhaul and can potentially relax the corresponding requirements regarding overall capacity, delay, and synchronization. This is also applicable to the concept of \ac{C-RAN} which is a potential implementation of \ac{CoMP} network. However, a study showing the feasibility of lower split options illustrates the preference of standardization in this regard \cite{3GPP_38_816}. 
    
    \item \ac{MIMO} enhancements in \ac{5G}, including multi-panel/\ac{TRP} operation distinguish \ac{5G} \ac{MIMO} from the \ac{LTE} \ac{MIMO} operation. Furthermore, improved (type II) codebook, flexible \ac{CSI} acquisition and reference signal design (including zero-power signals for interference measurement), and beam management for higher ($>$ 6GHz) bands promise significant boost in \ac{MIMO} performance \cite{3GPP_RP182863}. 
    

    

    \item Different \ac{5G} services are expected to have different coverage areas. This diversity of application requirements would be further pronounced in 6G. This can be intuitively interpreted as equivalent to cell edges in previous generations. Figure \ref{fig:Application Coverage} shows the preliminary simulation of coverage areas for different throughput requirements. The lack of uniform coverage motivates coordinated and cooperative communication to ensure a smooth user experience. 
    
    \item The lofty goals of \ac{5G} have motivated the exploration of higher frequency spectrum (\ac{mmWave} and \ac{THz}). However, they are susceptible to higher path loss and blockages. Technologies like beamforming and \ac{RIS} have been introduced to facilitate communication in this portion of the spectrum. Moreover, the use of macrodiversity offered by \ac{CoMP} has been experimentally shown to provide link and capacity improvement in the 73 GHz band \cite{maccartney2019millimeter}.
    
    
    
    \item Like \ac{4G}, \ac{5G} is also looking to incorporate unlicensed spectrum like 4G in its fold for improved network capacity, as evident from the presence of work item/study in Rel-16. \ac{5G} also enables cells to operate standalone in unlicensed spectrum, connected to a 5G core network \cite{3GPP_RP191575} \cite{3GPP_38_889}. Coordinated networks have the potential to improve resource utilization in the unlicensed and shared spectrum. 
    
    \item Related closely to this is the \ac{ATSSS} concept, enabling the simultaneous use of \ac{3GPP} and non-\ac{3GPP} access networks with the \ac{5G} core network. This is essentially one step beyond mere coexistence of different networks, leading to their convergence.
    

\end{itemize}

\section{\ac{GCoMP} Framework}
\label{sec:framework}
\begin{figure*}
	\centering
     \includegraphics[scale = 0.45]{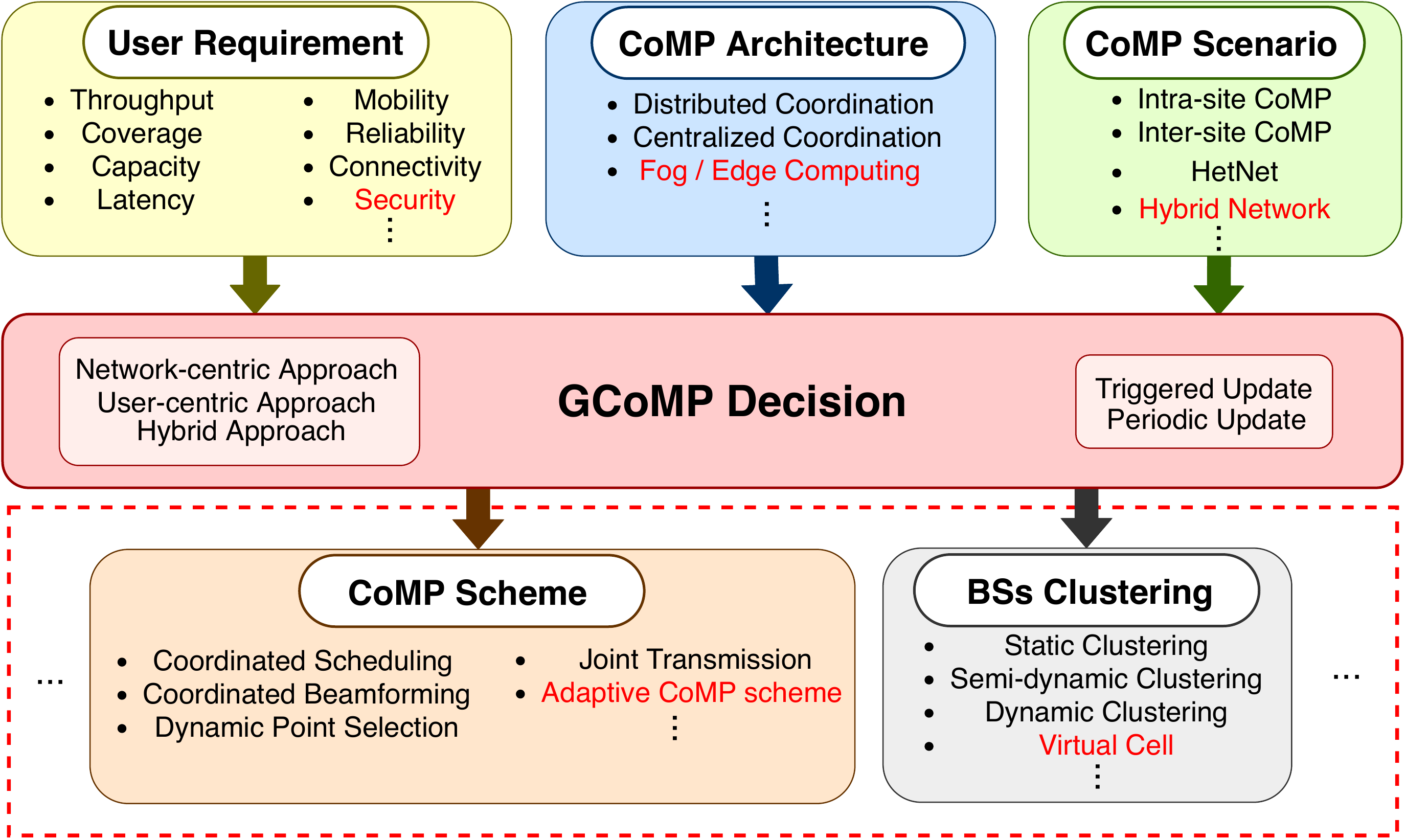}
     \caption{GCoMP conceptual framework. User requirements, CoMP architecture and scenarios serve as inputs to the decision mechanism. The outputs of this mechanism include (but are not limited to) selection of CoMP scheme and coordinating cluster. 
     }
     \label{fig:GCoMP Framework}
\end{figure*}

 The aforementioned \ac{5G} services require massive connectivity with high system throughput and improved spectral efficiency which imposes significant challenges for network design. We believe that the scope of \ac{CoMP} can be widened from mere interference mitigation to intelligent network resource utilization, helping achieve these diverse requirements. This section is dedicated to the description of the conceptual \ac{GCoMP} framework, illustrated in Fig. \ref{fig:GCoMP Framework}. The first row of elements represents the inputs to the \ac{GCoMP} decision mechanism. The decision making is the intermediate stage, followed by the outputs at the end. Here, it should be noted that while most options in the inputs/outputs are well-established, we have taken the liberty of identifying some additional ones, shown in red, that are either relatively new in general or at least recent to \ac{CoMP}.

\subsection{Inputs}
The input elements include user requirements, \ac{CoMP} architecture, and scenarios. The requirements are considered first since everything that follows revolves around them. These include reliability, throughput, capacity, mobility, and so on. \ac{CoMP} systems, also referred to as distributed \ac{MIMO} due to the presence of physically separated antenna locations, are capable of exploiting the spatial dimension to satisfy these requirements. For instance, if the multiple \acp{TP} are used for multiplexing, it increases the capacity of the system. On the other hand, reliability can be enhanced if different links are used to provide macrodiversity \cite{maladi}. In addition to the above-mentioned requirements, security is also a concern for mission-critical applications. Different \ac{PHY} and \ac{MAC} layer security algorithms can leverage the spatial diversity of \ac{CoMP} for providing security using techniques like directional modulation \cite{hafez2017secure}.    

Following requirements, the second input considered is the architecture. The conventional categories include centralized or distributed coordination. In the former, all administrative tasks are controlled through a central unit, while in the latter, one of the cooperating \acp{TP} acts as a master cell and performs all resource management and communication tasks. Here it is pertinent to mention the concept of centralized or \ac{C-RAN}, which has gained significant traction with operators due to its promise of reduced capital and operating expenditures. Despite its promise, one major challenge for \ac{C-RAN} is to balance the tradeoff between easier network management offered by centralized control and the increasingly strict backhaul bandwidth and latency requirements. This might be critical for use cases like \ac{V2X} communication. In light of this, recent works have proposed the utilization of fog/edge computing to provide intelligence to components of the network close to the \ac{UE} \cite{chen2018ultra}.


The third input element is \ac{CoMP} scenarios. \ac{3GPP} proposed three different \ac{CoMP} scenarios for both homogeneous and \acp{HetNet} \cite{3GPP_36_819}. The first scenario is homogeneous intra-site \ac{CoMP}, in which the coordination takes place between different \acp{TP} (sectors). Due to the colocation, there is no additional load on the backhaul. The second scenario is inter-site \ac{CoMP} which is also implemented on a homogeneous network. It uses high power \acp{RRH} to expand the coverage. The third scenario is implemented on \acp{HetNet} and utilizes low power \acp{RRH}. Inter-site \ac{CoMP} and \ac{HetNet} scenarios require high-speed backhaul links, like fiber, to make the connection between the macro cells and their respective \acp{RRH}. In line with \acp{HetNet}, another scenario that may be of interest is hybrid aerial-terrestrial networks. The wireless propagation channel characteristics of the air-to-ground channel are fairly different as compared to the conventional terrestrial channel, and can be taken advantage of to provide a better \ac{QoS} to the \acp{UE} \cite{solaija2019hybrid}. This can be extended to incorporate the non-terrestrial network or satellite communication scenarios, aimed at improving network coverage \cite{3GPP_SP180326}. The logical next step to exploiting the variation in the propagation environment is the capability of modifying the environment itself to improve the coverage and user experience. \ac{RIS} is a technology that promises exactly that by selectively modifying the incident signal's properties, such as phase, amplitude, and polarization \cite{saad2019vision}.

Here it is important to categorize the nature of the above-mentioned inputs in terms of their dynamicity. User requirements are expected to change depending on the application being used. The architecture is static, while the scenario might change, particularly in terms of intra/inter-site coordination.


\subsection{Decision Making}
The \ac{GCoMP} decision-making process evaluates the above-mentioned input elements, network constraints, and channel conditions to make informed decisions regarding the appropriate resource allocations, namely, selection of the best suited \ac{CoMP} scheme and coordination cluster. The approaches for this process can be categorized into \textit {user-centric}, \textit{network-centric}, or \textit{hybrid}. The user-centric approach would make decisions on a per-user basis, targeted at fulfilling that particular user's requirements. The network-centric decision making, on the other hand, places more emphasis on simplifying the implementation from the network perspective, including the architecture and overhead while trying to optimize the performance of all connected users. The overhead includes information (data and \ac{CSI}) sharing between the nodes and processing of the information necessary for the said decision-making. The hybrid approach provides a tradeoff between both the above-mentioned methods by optimizing the decisions for a group of users while keeping the network overhead bearable. 

\begin{figure}[t!]
    \centering
    \includegraphics[scale=0.54]{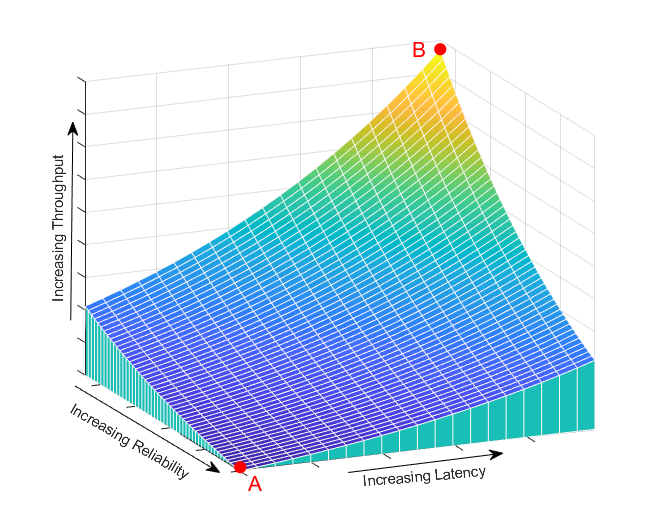}    \caption{A sketch illustrating the tradeoff between latency, reliability and throughput (inspired from \cite{soret2014fundamental})}
    \label{fig:Trade off}
\end{figure}

Given that the future networks are expected to have an exceedingly high number of devices, the available resources might be insufficient to serve them. There are two ways to go about solving this problem. The network can prioritize the users depending on the required \ac{QoS} where the prioritization information associated with \ac{5QI} can be used for this purpose \cite{3GPP_23_501}. Alternatively, it can try to optimize the varying and occasionally competing user requirements. An example of this is the tradeoff between latency, reliability, and throughput, as illustrated in Fig. \ref{fig:Trade off}. While it is possible to optimize any two of the requirements, it comes at the cost of the third \cite{soret2014fundamental}. This is visible for point A in Fig. \ref{fig:Trade off} where reliability and latency are optimized, but throughput is compromised. Point B, on the other hand, provides the opposite. In such scenarios, a single optimum solution is not possible. Rather, there is a set of (possibly infinite) Pareto-optimal solutions where improving one objective would lead to degradation in the other(s) \cite{bjornson2014multiobjective}. Network slicing is capable of complementing the optimization problem for resource allocation by providing portions of the network resources for different use cases. The framework governing network slicing in \ac{5G} is quite generic and allows the possibility of assigning multiple \acp{NSI} with different characteristics to a certain communication service \cite{3GPP_28_801}. This can be leveraged to support the otherwise competing user requirements like the ones mentioned above. 


Since these decisions are dependent upon variable parameters such as user requirements and spatio-temporal traffic patterns, they need dynamic updates. These updates can either be \textit{periodic} or \textit{triggered}. As the name suggests, the former analyzes the network situation repeatedly after a fixed interval and revisits its earlier decisions, while the latter can be set off by certain nodes and/or conditions. This may include variation in the backhaul availability or traffic patterns.  









\subsection{Outputs}

The first output of the framework is the selection of the appropriate \ac{CoMP} scheme. These schemes have different backhaul requirements and provide varying benefits. Since \ac{C-JT} performs joint beamforming, it requires backhaul links with high capacity and low latency as well as strict synchronization among coordinated \acp{TP}. \ac{NC-JT}, on the other hand, provides a complexity-performance tradeoff by removing the burden of joint precoding and strict synchronization while still providing significant gains as compared to other schemes such as \ac{CS}, \ac{CB}, and \ac{DPS} \cite{3GPP_36_741}. Therefore, the \ac{GCoMP} decision needs to consider each \ac{UE}'s requirements and weigh it against the backhaul bandwidth before making a decision. One interesting approach presented in \cite{su2018interference}, where an attempt to mitigate the backhaul delay caused during \ac{UE} data sharing is made by adaptively changing the transmission modes (from \ac{CS}/\ac{CB} to \ac{JT}).   

The second output identified for this framework is the decision about the coordination cluster, which comprises of the \acp{TP} that are supposed to coordinate with each other. In literature, there are three main types of clustering. \textit{Static} clustering, which is primarily based on topology and does not vary according to the nodes or \acp{UE}, provides limited performance gains. \textit{Semi-dynamic} clustering - an enhanced version of the former - where more than one static clustering patterns are set up and \acp{UE} can select the most suitable cluster, leads to an increase in both complexity and performance. \textit{Dynamic} clustering responds to network and \ac{UE} mobility changes and reduces inter-cluster interference by updating the clusters dynamically. To identify the coordinated \acp{TP} per cluster, a set of solutions is proposed in \cite{3GPP_36_742} taking into account real operating conditions such as connectivity and network layout. One of the solutions is to adapt the \acp{CA} depending on the spatial distribution of the \acp{UE} in order to avoid concentrations of \acp{UE} on inter-\ac{CA} borders. Another solution is the use of layered \acp{CA} where the borders between adjacent \acp{CA} are covered by an overlaying \ac{CA}. Indeed, a coordinated \ac{TP} can be part of different \acp{CA} and partitioning of scheduler resources between the \acp{CA} is needed which might cause some peak \ac{UE} throughput limitations. Therefore, \ac{CA} layers should be activated only when needed. 

In addition to the clustering approach, there is the concept of virtual cell \cite{chen2018ultra}, where each such cell is occupied by a single \ac{UE}. This \ac{UE} is served by multiple cooperating \acp{TP} leveraging the concept of network slicing where different logical slices of the network are used to facilitate different \acp{UE}. 


\section{Use Cases}
\label{sec:use cases}

\begin{figure*}[t!]
    \centering
\includegraphics[width=\textwidth]{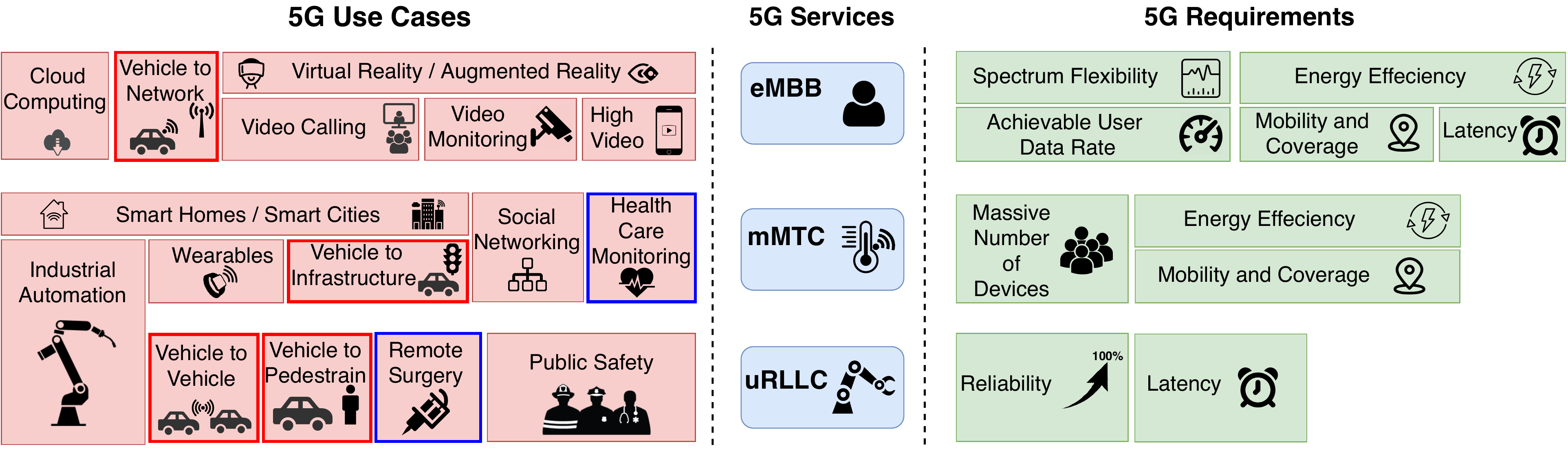}
    \caption{Selected 5G use cases, services and requirements}
    \label{fig:use cases}
\end{figure*}

\ac{5G} networks are envisaged to support a large number of verticals, ranging from healthcare to industrial automation, intelligent transport to immersive entertainment, and so on. The variety of targeted verticals leads to diverse requirements, which are grouped under different \ac{5G} services. Figure \ref{fig:use cases} illustrates the services and concerned requirements for some selected use cases discussed under \ac{5G} \cite{5Gservices&usecases}. These include virtual/augmented reality and high definition video as examples of the \ac{eMBB} service,  public safety networks, which form part of the \ac{uRLLC} fabric and smart homes and wearable technologies under the category of \ac{mMTC} service (the latter is yet to be properly addressed in the standard even though Rel-16 provides some enhancements for NB-IoT and \ac{eMTC} \cite{3GPP_21_916}). 
Additionally, we have certain cross-service use cases such as eHealth and \ac{V2X} communications. The following passages provide a brief overview of these use cases, their requirements, and how they may be facilitated under the \ac{GCoMP} framework. 

\subsection{V2X Communication: Overview and Requirements}
\label{subsec:V2X}
Autonomous vehicles or \acp{ITS} are driven by two primary motivations, firstly to reduce traffic accidents and consequent fatalities or injuries and secondly, to reduce traffic congestion. The efficient commute also has the advantages of lowered fossil fuel consumption and reduced carbon dioxide emissions. Ubiquitous communication between vehicles, on-board sensors, and the environment is imperative to ensure the smooth operation of \ac{ITS}. \ac{V2X} communication is the umbrella term used for the vehicles' communication with other vehicles, infrastructure, network, and pedestrians. 

Considering the various aspects and scenarios of \ac{V2X} communications, the following requirements can be listed:
\begin{itemize}
    \item Provision of \textbf{ultra-reliable} and \textbf{low-latency} communication to ensure proper operation of \acp{ITS}.
    \item \textbf{Mobility} is a key characteristic of \acp{ITS} which causes rapid channel variation, making the communication very challenging. Providing ubiquitous connectivity to these \acp{UE} is of paramount importance.
    \item Since autonomous vehicles rely heavily on \ac{V2X} communication, they are susceptible to various forms of attacks to the communication that may disrupt transport or even hurt people. Identifying these \textbf{security} concerns and coming with appropriate solutions is a relatively new challenge that requires significant research efforts. 
\end{itemize}

As mentioned earlier, user prioritization and \ac{QoS} information can ease the burden of \ac{GCoMP} decision-making. Therefore the availability of different aspects of \ac{V2X} communications under \ac{5QI} values 83-86 in \cite{3GPP_23_501} can be leveraged for this purpose.


\subsection{eHealth: Overview and Requirements}
\label{subsec:eHealth}
eHealth refers to the use of information and communication technologies for improved healthcare. One of the major reasons for this push from health organizations is the lack of qualified professionals in the developing world. Remote services like monitoring, consultation, and surgery are primarily being focused on under this \ac{5G} vertical.

The components of this \ac{5G} vertical include monitoring of patient information such as blood glucose or blood pressure using a wide variety of sensors, online consultation of patients by medical practitioners, and arguably the most hyped use case, remote surgery. The following requirements can be extracted from these scenarios \cite{ullah20195g}:
\begin{itemize}
    \item \textbf{Ultra-reliable} and \textbf{low-latency} communication for the control of surgical robots is needed.
    \item \textbf{High throughput} for video monitoring, consultation, or remote surgery is required.
    \item There can be a wide variety of sensors under use, some of which might be implanted \textit{in vivo}. In such a case \textbf{energy conservation} becomes critical.
\end{itemize}

It is interesting to point out that depending on the above-mentioned requirements, this particular use case seems to lie on the intersection of \ac{eMBB} and \ac{uRLLC} services which maps neatly to the standardized \ac{5QI} value of 80 \cite{3GPP_23_501}.

\subsection{GCoMP Decision Flow}
\begin{figure*}
	\centering
     \includegraphics[scale = 0.9]{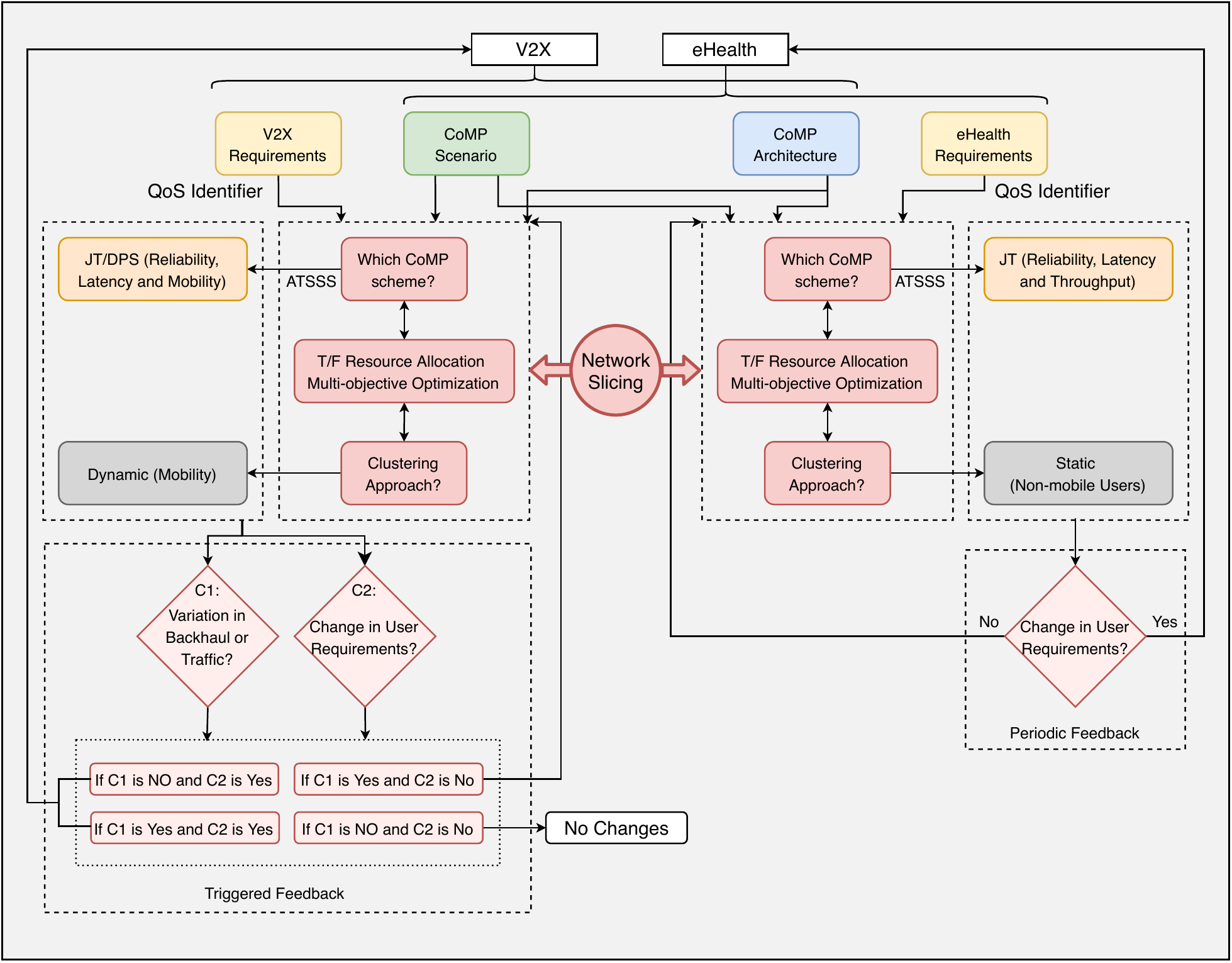}
     \caption{Flow of the GCoMP decision mechanism for the V2X and eHealth use cases. 
     }
     \label{fig:Flowchart}
\end{figure*}

This section describes the working of \ac{GCoMP} mechanism for instances of the use cases defined above, as shown in Fig. \ref{fig:Flowchart}. It can be seen that for both use cases/applications the requirements are derived using \ac{5QI} \cite{3GPP_23_501} while the \ac{CoMP} architecture and scenarios are considered as inputs from the network. For easier understanding, we do not put any restriction on either at present and allow the use of different network slices for both applications. Depending on if there is a single or multiple users belonging to the same application requirements (and consequently the concerned network slice), the decision-making approach can be considered as user-centric or hybrid, respectively.

This approach allows the independent selection of clustering and \ac{CoMP} scheme for both use cases. However, since both applications need high reliability and low latency, the use of \ac{JT} seems suitable since it allows the transmission of multiple copies of the same signal to the \acp{UE}. \ac{DPS} can also be considered for \ac{V2X} due to the mobility-induced handovers. \ac{ATSSS} is also capable of supporting the mobility (handover) and throughput requirements with its switching and splitting aspects, respectively \cite{men2019MPTCP}.

In terms of clustering, a dynamic approach is needed for the \ac{V2X} case while static clustering can suffice for eHealth. As mentioned earlier, the network itself is dynamic and needs to be taken into consideration in our framework. Owing to the increased spatio-temporal traffic variation in the \ac{V2X} case, it can use the triggered feedback approach, where the decision making can be updated depending on both user requirements and network variation. eHealth, on the other hand, is not expected to observe much variation in that regard, so a periodic update should be able to suffice.

The authors would like to reiterate that this is a very simple example to illustrate the \ac{GCoMP} framework. It is possible to incorporate more variables and improve the decision flow to either optimize specific scenarios or generalize it to cater to a wider variety of applications.

\subsection{CoMP-based Solutions}
Described below are the different approaches used under the context of coordinated networks for the requirements pointed out in Sections \ref{subsec:V2X} and \ref{subsec:eHealth}:

\subsubsection{Reliability and Latency}
Out of the different services of \ac{5G}, \ac{uRLLC} is arguably the toughest challenge owing to the targeted reliability with strict latency bounds. There are genrrally two approaches to address the \ac{uRLLC} requirements, increasing reliability of one-shot transmission or to lower the latency between retransmissions. \ac{CoMP}, with its \ac{JT} approach, can provide different versions of the transmitted signal at the receiver at the same time reducing the necessity of retransmission and addressing the latency constraint. Properly combining the received copies of the signal can improve the \ac{SINR} performance and hence reliability of the system. 
Macrodiversty is an approach to provide multiple paths for the communication of the same signal, targeted to exploit the variation of path loss and large scale fading in the different paths. An interesting idea related to this is to utilize hybrid aerial-terrestrial networks, which provide additional diversity in the wireless link owing to the different propagation characteristics of the air-to-ground channels \cite{solaija2019hybrid}. In these networks, the aerial \acp{TP} provide a much higher probability of line of sight and reduced shadowing, resulting in improved reliability of communication.
An alternative to this is the packet duplication approach supported in Rel-15 \cite{3GPP_21_915} which is a higher (\ac{PDCP}) layer complement of the \ac{PHY} layer diversity techniques \cite{rao2018packet}.

\subsubsection{Mobility}
Due to the inherent mobility of \acp{TRP} in \ac{V2X} communications, handovers are a major concern. This becomes critical especially when talking about cell densification, small cells, and the strict reliability and latency requirements. Additionally, the signaling overhead associated with each handover and possible service interruption are undesirable from the perspective of network and \ac{UE}. To ensure continuous connectivity, \textit{dual connectivity} based solution was considered (though not eventually standardized) in addition to \ac{MBB} handover technique \cite{park2018handover}. \ac{CoMP} or \ac{C-RAN} can be a form of implementation for multi-connectivity. Additionally, \ac{CoMP} also reduces the number of handovers as long as the \ac{UE} is within its coordinating cluster. A \ac{CoMP} scheme like \ac{DPS} seems particularly suitable for mobile \acp{UE}, owing to the similarity in nature of handover and \ac{DPS} concepts since both revolve around the dynamic selection of best-suited \ac{TP}. Along the same lines, the switching aspect of \ac{ATSSS} is capable of supporting smoother handovers by leveraging \ac{MP-TCP} \cite{MPTCP_IETF}. 

\subsubsection{Security}
The spatial diversity offered by \ac{CoMP} can be leveraged to establish secure wireless communication links between legitimate \acp{UE} if it is merged with \ac{PHY} layer techniques. The geographically separated \acp{TP} can be used to transmit the signal such that data is only decodable at the intersection of their transmission beams \cite{hafez2017secure}, providing location-based security against eavesdropping. Jamming is another threat to communication. These attacks include transmission of noise or noise-like signals to interrupt the legitimate transmission. The spatial diversity offered by the geographically separated \acp{TP} may be utilized to combat such attacks.

\subsubsection{Throughput}
As mentioned earlier, the spatial diversity offered by \ac{CoMP} systems can be used in various ways. \ac{JT}-\ac{CoMP} promises significant gains in terms of network capacity and \ac{UE} throughput by combining signals from different \acp{TP} either coherently or non-coherently. Coherent \ac{JT} is capable of providing higher throughput as compared to its non-coherent counterpart since it uses a joint precoding procedure while the latter focuses on improving the received signal strength \cite{qamar2017comprehensive}. Multi-\ac{TRP} \ac{MIMO} utilizing beamforming at \ac{mmWave} bands promises increased data rates. Furthermore, this requirement can also leverage the \ac{MP-TCP} and underlying \ac{ATSSS} concept to split the traffic over multiple access networks, resulting in improved throughput for the user \cite{men2019MPTCP}.

\subsubsection{Energy Conservation}
Energy conservation, particularly for the \acp{UE}, is a major concern in wireless networks. It especially holds true for sensors that are not easily accessible, medical implants being a perfect example. In the case of multiple devices that need to communicate, it is possible to consider \ac{DPS} with energy conservation as a goal. A similar idea in drone-based disaster recovery scenario is proposed in \cite{mezghani2019disaster} where the uplink \ac{TP} is selected from the \acp{UE} while taking into consideration the remaining battery of the \acp{UE}.

\section{Challenges and Future Directions}
\label{sec:challenges}
There are considerable challenges that need to be overcome in order to make \ac{GCoMP} a reality. Some of these issues are identified below:

\subsection{Optimization}
Owing to the diversity of future wireless networks both in terms of user/application requirements and device/node capabilities, optimized resource allocation is going to become even more challenging. Multi-objective optimization is, therefore, going to be imperative. This also includes the need for improved network slicing capabilities which will be necessary to support future applications.  

\subsection{Inter-RAT Coordination}
The emergence of \ac{ATSSS} signals the convergence of different \ac{RATs}. Furthermore, the upcoming amendment of the Wi-fi standard, i.e., IEEE 802.11be has introduced multi-\ac{AP} coordination concept which is similar to \ac{CoMP}. This illustrates the importance of and need for coordination within and between different \ac{RATs}. However, the interface and exact mechanisms to carry out these tasks need to be defined and standardized (current Wi-Fi activities related to multi-AP coordination are in infancy stages \cite{khorov2020current}).

\subsection{Next Generation Networks}
Since 6G is still an abstract concept at this point, with most works considering it an extension of the original \ac{5G} requirements, we have not delved much upon the issue. However, from recent activities it is evident that secure, smart radio environments (using \acp{RIS}) and intelligent networks are going to be an imperative part of future communications \cite{saad2019vision} \cite{dang2020should}. The fact that \ac{GCoMP} concept itself originates from the need for an intelligent and proactive system highlights the possibility of pushing it even further with these enablers. 

\subsection{Network Overhead Considerations}
Some guidelines that need to be considered for a system as extensive and comprehensive as \ac{GCoMP} can be borrowed from \cite{3GPP_36_742}. This includes analyzing the convergence of the CoMP function on an appropriate time-scale, impact on \ac{eNB} complexity, additional signaling, and any impact of the network configuration.

\subsection {Spatial Diversity Tradeoffs} 
The spatial diversity afforded by the geographical separation between \acp{TP} can be utilized to provide capacity, security, and reliability gains. However, all of these are competing objectives which means one can only be achieved if the others are waived. Optimizing these tradeoffs remains a challenge.  

\subsection{Adaptive CoMP Schemes}
The concept of \ac{GCoMP} includes selecting the best \ac{CoMP} scheme for a given \ac{UE} taking into account the network architecture and scenario. However, the capability of a network to support multiple \ac{CoMP} schemes simultaneously still needs to be studied and its advantages weighed against the potential costs. 

\subsection{CoMP Triggering}
Traditionally, cell edge is defined by static thresholds of parameters like \ac{SINR} and this is where \ac{CoMP} was applied. However, with diverse services, these thresholds are going to vary with requirements. In such a situation a better triggering mechanism is required.
 
\subsection{Traditional CoMP Issues}
Some of the problems that have limited the practice of \ac{CoMP} include imperfect \ac{CSI}, insufficient backhaul, and clock synchronization \cite{qamar2017comprehensive}. Generally speaking, these issues degrade the throughput, reduce the number of connected \acp{UE}, cause inter-symbol or inter-carrier interference, and so on. Here it is necessary to mention that these impairments would affect the various \ac{CoMP} schemes differently, which needs to be taken into consideration when selecting a scheme for given requirements.

\section{Conclusion}
\label{sec:conclusion}
\ac{5G} embodies a significant evolution in wireless communication by incorporating services like \ac{eMBB}, \ac{mMTC}, and \ac{uRLLC}. However, the fulfillment of these services is quite challenging owing to the diversity in their requirements. Driven by the realization that presently available techniques are unable to support the lofty targets set for \ac{5G} unless a significant effort to build a coordinated network is made, we have proposed the generalization of \ac{CoMP} concept. The aim is to expand the scope of \ac{CoMP} from mere interference management at cell edges to enhancing the throughput, decreasing latency, increasing reliability, improving coverage, and providing seamless connectivity to \acp{UE} with varying requirements. To this end, a generalized \ac{CoMP} framework has been discussed in this paper, which we believe will prove to be a stepping stone towards the realization of fully coordinated \ac{5G} and beyond networks.

\section{Acknowledgment}
This work was supported in part by the \ac{TUBITAK} under Grant No. 116E078. The work of M. I. Saglam was supported by \ac{TUBITAK} under Grant No. 3171084. The authors thank Halise T{\"u}rkmen for her helpful suggestions regarding this manuscript.


\bibliographystyle{IEEEtran}

\begin{IEEEbiographynophoto}{Muhammad Sohaib J. Solaija}[S'16] (solaija@ieee.org) received his B.E and M.Sc degrees in electrical engineering from National University of Science and Technology, Islamabad, Pakistan in 2014 and 2017, respectively. Currently he is pursuing his Ph.D. degree at Istanbul Medipol University, Turkey. His research focuses on interference modelling and coordinated multipoint (CoMP) implementation for 5G and beyond wireless systems.
\end{IEEEbiographynophoto}

\begin{IEEEbiographynophoto}{Hanadi Salman}[S'18]~received her B.S. degree from An-Najah National University, Nablus, Palestine in 2017. She is currently pursuing her Ph.D. degree at Istanbul Medipol University, Turkey. Her research focuses on interference modeling, resource scheduling, and coordinated multipoint (CoMP) for 5G and beyond wireless systems.
\end{IEEEbiographynophoto}

\begin{IEEEbiographynophoto}{Abuu B. Kihero} received his B.S. degree in electronics engineering from Gebze Technical University, Kocaeli, Turkey in 2015, and M.S. degree  in electrical, electronics and cyber systems from Istanbul Medipol University, Istanbul, Turkey in 2018. He is currently pursuing his Ph.D degree in wireless communication at the latter. His research field of interest includes wireless channel modeling and emulation, interference modeling and coordinated multipoint (CoMP) for 5G and beyond wireless systems.
\end{IEEEbiographynophoto}

\begin{IEEEbiographynophoto}{Mehmet Izzet Sa\u{g}lam} received his B.S. degree in electrical and electronics engineering from \c{C}ukurova University in 2001, and his M.S. and PhD degrees in communication from Istanbul Technical University, Istanbul, Turkey, in 2003 and 2018, respectively. He is a researcher and 3GPP RAN WG2 delegate at Turkcell R\&D. His current research interests are on radio resource management of wireless systems. 
\end{IEEEbiographynophoto}

\begin{IEEEbiographynophoto}{H{\"u}seyin Arslan}[S'95--M'98--SM'04--F'15] received his B.S. degree in electrical and electronics engineering from Middle East Technical University in 1992, and his M.S. and PhD degrees in electrical engineering from Southern Methodist University, Dallas, Texas, in 1994 and 1998, respectively. He is a professor of electrical engineering at the University of South Florida and the Dean of the School of Engineering and Natural Sciences at Istanbul Medipol University, Turkey. His current research interests are on physical layer security, mmWave communications, small cells, multi-carrier wireless technologies, co-existence issues on heterogeneous networks, aeronautical (high altitude platform) communications and \textit{in vivo} channel modeling, and system design. 
\end{IEEEbiographynophoto}

\end{document}